\documentclass[traditabstract]{aa}

\usepackage{graphicx}
\usepackage{multirow}
\usepackage{natbib}
\usepackage{adjustbox}
\usepackage{txfonts}
\usepackage{lipsum} 
\usepackage{mathrsfs}
\usepackage{breqn}
\usepackage[colorlinks = true,linkcolor = blue,urlcolor = blue,citecolor = blue]{hyperref}
\usepackage{xcolor}
\definecolor{yellowgray}{rgb}{0.90, 0.90, 0.2}
\definecolor{bluegray}{rgb}{0.20, 0.60, 0.80}
\definecolor{palered}{rgb}{0.99, 0.40, 0.5}
\definecolor{darkgray}{rgb}{0.35, 0.35, 0.35}
\definecolor{darkgrayb}{rgb}{0.75, 0.75, 0.75}
\definecolor{palegray}{rgb}{0.96, 0.96, 0.96}

\begin{document} 

\title{A study of the capabilities for inferring atmospheric information from high-spatial-resolution simulations}
  \titlerunning{Capabilities for inferring atmospheric information from high-spatial-resolution simulations}
   \author{C. Quintero Noda \inst{1,2}\thanks{Corresponding author}  \and E. Khomenko\inst{1,2} \and M. Collados\inst{1,2} \and B. Ruiz Cobo\inst{1,2} \and R. Gafeira\inst{3} \and  N. Vitas\inst{1,2}  \and M. Rempel\inst{4} \\ R. J.~Campbell\inst{5} \and A. Pastor Yabar\inst{6} \and H. Uitenbroek\inst{7} \and D. Orozco Su\'{a}rez\inst{8}
             }
   \institute{ Instituto de Astrof\'isica de Canarias, E-38200, La Laguna, Tenerife, Spain.\\
              \email{carlos.quintero@iac.es}     
\and  Departamento de Astrof\'isica, Univ. de La Laguna, La Laguna, Tenerife, E-38205, Spain
\and Instituto de Astrof\'{i}sica e Ci\^{e}ncias do Espa\c co, Departamento de Física, Universidade de Coimbra, Rua do Observat\'{o}rio s/n, 3040-004 Coimbra, Portugal
\and High Altitude Observatory, NCAR, P.O. Box 3000, Boulder, CO 80307, USA
\and Astrophysics Research Centre, School of Mathematics and Physics, Queen’s University Belfast, Belfast, BT7 1NN, Northern Ireland, U.K.
\and Institute for Solar Physics, Department of Astronomy, Stockholm University, AlbaNova University Centre, 10691 Stockholm, Sweden
\and National Solar Observatory, University of Colorado Boulder, 3665 Discovery Drive, Boulder, CO 80303, USA
\and    Instituto de Astrof\'isica de Andaluc\'ia (CSIC), Apdo. de Correos 3004, E-18080 Granada, Spain\\
             }
   \date{Received 25 March 2021 ; accepted 28 January 2022  }


 
\abstract{In this work, we study the accuracy that can be achieved when inferring the atmospheric information from realistic numerical magneto-hydrodynamic simulations that reproduce the spatial resolution we will obtain with future observations made by the 4m class telescopes DKIST and EST. We first study multiple inversion configurations using the SIR code and the \ion{Fe}{i} transitions at 630~nm  until we obtain minor differences between the input and the inferred atmosphere in a wide range of heights. Also, we examine how the inversion accuracy depends on the noise level of the Stokes profiles. The results indicate that when the majority of the inverted pixels come from strongly magnetised areas, there are almost no restrictions in terms of the noise, obtaining good results for noise amplitudes up to 1$\times10^{-3}$ of $I_c$. At the same time, the situation is different for observations where the dominant magnetic structures are weak, and noise restraints are more demanding. Moreover, we find that the accuracy of the fits is almost the same as that obtained without noise when the noise levels are on the order of 1$\times10^{-4}$of $I_c$. We, therefore, advise aiming for noise values on the order of or lower than 5$\times10^{-4}$ of $I_c$ if observers seek reliable interpretations of the results for the magnetic field vector reliably. We expect those noise levels to be achievable by next-generation 4m class telescopes thanks to an optimised polarisation calibration and the large collecting area of the primary mirror.}


\maketitle

\section{Introduction}

Understanding the physics of solar phenomena implies the inference of information about their thermal and magnetic properties. In the case of the magnetic field, there are multiple ways to infer that information, from basic techniques such as deriving the field strength from the separation between the Stokes $I$ Zeeman components \citep[e.g.,][]{2004ASSL..307.....L}, comparing the derivative of the Stokes $I$ profile with Stokes $V$ with the so-called weak-field approximation (see \cite{2017SSRv..210..109D} or \cite{2018ApJ...866...89C} for a recent example), to more elaborate techniques such as the so-called inversion codes \citep[see][ for an introduction]{2016LRSP...13....4D}. Regarding the estimation of the magnetic properties of quiet Sun regions using inversion techniques, there were multiple works in the early 2000s. Those works used observations made with ground-based telescopes, for instance, with the Tenerife Infrared Spectropolarimeter \citep{Collados2007} installed at the Vacuum Tower Telescope \citep{vonderLuhe1998} and the Advanced Stokes Polarimeter \citep{Elmore1992} installed in the Dunn Solar Telescope. The observations usually recorded the visible \ion{Fe}{i} line pair at 630~nm or the infrared \ion{Fe}{i} lines at 1.56 microns. The results of their analysis were published in multiple publications \citep[see, among others, ][]{Lites2002,Khomenko2003,Lites2004,SanchezAlmeida2004,DominguezCerdena2006}.

After that period, the solar community saw an exponential increase in the number of works dedicated to elucidating the magnetic properties of the quiet Sun, mainly because the Hinode mission \citep{2007SoPh..243....3K} was launched in 2006, in a period of minimum solar activity. The spacecraft offered high-stability observations \citep{2008SoPh..249..221S} over long periods thanks to the excellent performance of the Solar Optical Telescope \citep{2008SoPh..249..167T}. This stability allowed multiple types of quiet Sun observations to be made with short and long exposure times, over small and large fields of view, and at the disk centre and the limb and so on using the spectropolarimeter \citep{Lites2013} installed on the optical telescope. Later, those data were complemented with higher spatial resolution observations taken with the Imaging Magnetograph eXperiment \citep{2011SoPh..268...57M} on board the Sunrise balloon \citep{2010ApJ...723L.127S}, which was launched for the first time in 2009. There is a long list of publications from quiet Sun observations made using the Hinode satellite and the Sunrise balloon, and so with the best intention of not leaving any critical work behind, we cite the recent review of \cite{BellotRubio2019}, where those works are gathered and thoroughly explained. However, we cite a few particular publications that we consider relevant to the present article. One of the early works from the Hinode era was that of \cite{2008ApJ...672.1237L}, where the authors provided statistics regarding the properties of the magnetic field in the quiet Sun. This was followed by a plethora of related publications, with some works building on the dynamics of the quiet Sun magnetic fields \citep[for instance,][]{2007ApJ...666L.137C,2009ApJ...700.1391M,Danilovic2010}. 

In parallel, we also saw a rise in the number of works focusing on the possible limitations of studying weak polarisation signals that are dominated by noise using synthetic data generated from semi-empirical atmospheres as well as from realistic 3D magneto-hydrodynamic (MHD) simulations \citep[among others,]{2011A&A...527A..29B,2012A&A...543A..34D,2016A&A...593A..93D,2017ApJ...842...37B,2019A&A...630A.133M,2021A&A...654A..11C}. Those works tested the ability to retrieve the original atmosphere used as input for the synthesis of the Stokes profiles under different conditions of spatial and spectral degradation, or noise. We believe these works are an excellent reference for observers that aim to, for example, prepare a proposal for observations, and we want to expand their work further. We aim to reproduce observations that will soon be made by next-generation 4m class telescopes such as DKIST \citep{2020SoPh..295..172R} and EST \citep{2022A&A...666A..21Q}. To this end, we use state-of-the-art high-spatial-resolution numerical simulations with a pixel element with the size expected at the diffraction limit of those telescopes. Also, we aim to set the basis for highly accurate inversions where the user of inversion codes seeks to infer the stratification of the atmospheric parameters at different layers, obtaining precise fits of complex Stokes profiles through the use of gradients along the line of sight (LOS) on temperature, velocity, and the magnetic field vector. Thus, we aim to provide an inversion strategy from which future observers at the aforementioned next-generation telescopes can start from. Moreover, we are curious to see how the inversion technique behaves in such complex numerical scenarios when different noise levels are included and try to set a reference noise level that observers would need to aim for depending on their research interests.

\begin{figure}
\begin{center} 
   \includegraphics[trim=0 0 0 0,width=8.5cm]{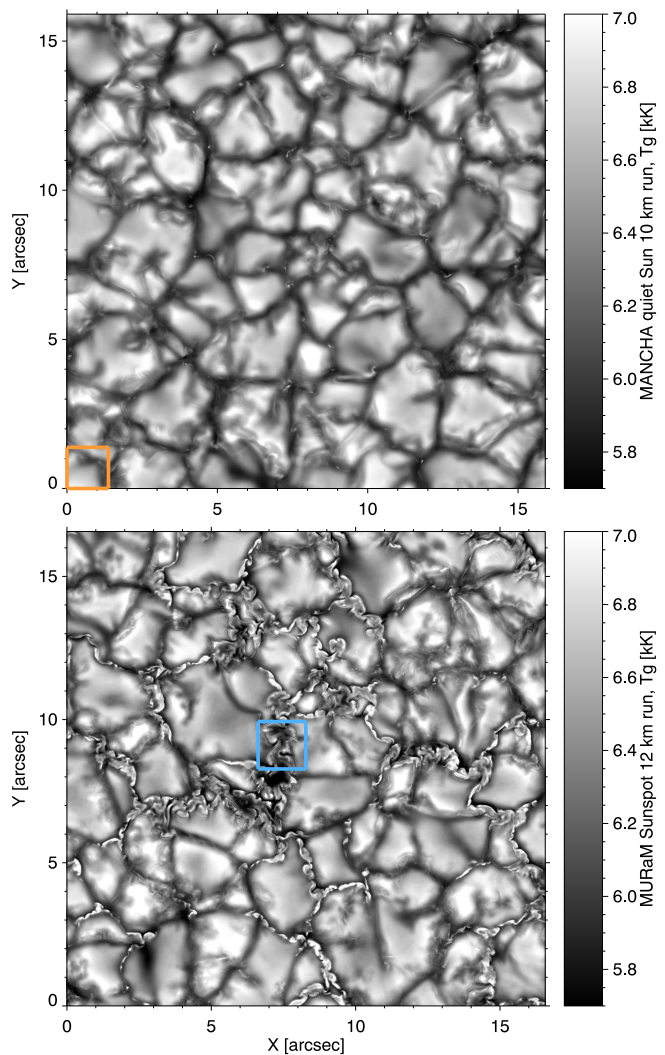}
 \vspace{-0.15cm}
 \caption{Spatial distribution of the temperature where the continuum optical depth at 500~nm is unity. The top panel shows the results from a quiet-Sun run with 10 km spatial resolution computed with the MANCHA code. The bottom panel corresponds to a cut of the original Sunspot run with a 12 km spatial resolution generated with the MURaM code. The regions highlighted with coloured boxes designate the areas of interest on which we focus in this work.}
 \label{FOV}
 \end{center}
\end{figure}

\section{Data and methodology}\label{Method}

\begin{figure*}
\begin{center} 
   \includegraphics[trim=0 0 0 0,width=18.2cm]{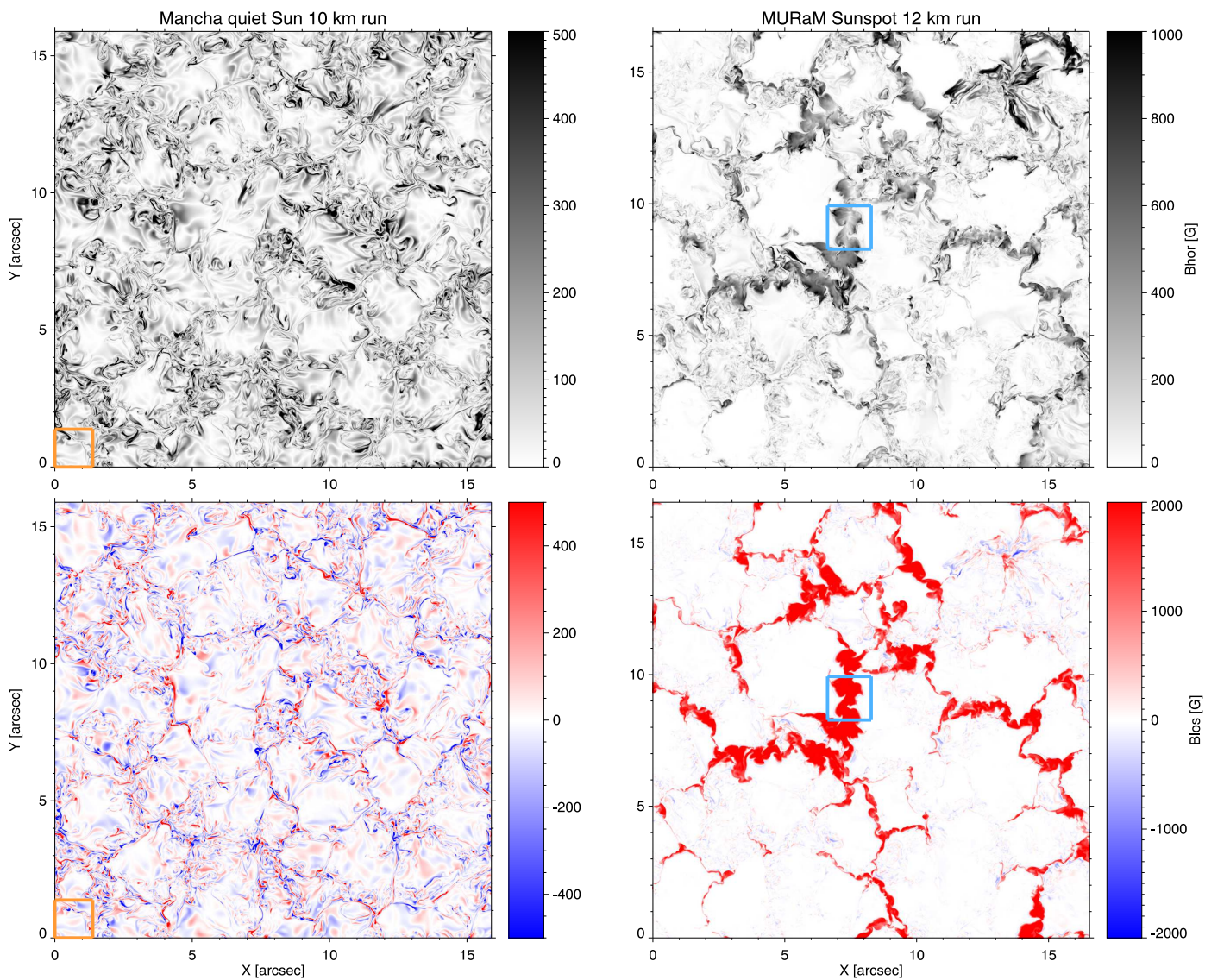}
 \vspace{-0.10cm}
 \caption{Spatial distribution of the horizontal (top) and longitudinal (bottom) component of the magnetic field where log~$\tau=0$. The left panel corresponds to a quiet Sun run with 10~km spatial resolution generated with the MANCHA code, while the right panel shows a cut of the original Sunspot run with 12~km spatial resolution computed with the MURaM code.}
 \label{B_2D}
 \end{center}
\end{figure*}

We use two different 3D snapshots in the present study. The first was computed with the MANCHA code \citep[][]{2006ApJ...653..739K,2010ApJ...719..357F,2018A&A...615A..67G} with the configuration described in \cite{2017A&A...604A..66K} and \cite{2018A&A...615A..67G}. The realistic convection simulation snapshot used here is a simulation run initiated as purely hydrodynamic convection in a box spanning from approximately 1~Mm below the photosphere to 600~km above it. The pixel size in the horizontal and vertical directions is 10 and 7~km, respectively, while the snapshot has a squared field of view (FOV) of 15.88$\times$15.88~arcsec$^{2}$ (a total of 1152$\times$1152~pix$^{2}$). Once the convection has reached a stationary state, an initial magnetic field was seeded by `switching on' the Biermann battery term in the induction equation \citep{2017A&A...604A..66K}. Its continuous action, together with the dynamo amplification, provided the spatially averaged magnetisation of the model of about $\langle |B| \rangle \approx 10^2$ G at the surface where the continuum optical depth at 500~nm is unity after a few hours of solar time. The snapshot we use in this work corresponds to the saturated dynamo regime.

The second snapshot corresponds to the `Sunspot' run from \cite{2012ApJ...750...62R} computed with the MURaM code \citep{2005A&A...429..335V}. The pixel size in the horizontal and vertical domains is 12 and 8~km, respectively. We focused only on the bottom left corner of the original field of view. This small subset contains 1000$\times$1000~pix$^{2}$ and has a total size of 16.55$\times$16.55~arcsec$^{2}$. 

The original atmospheres, computed on a regular vertical geometrical grid, were transformed to a 3D cube where the vertical domain corresponds to the continuum optical depth at 500~nm (log~$\tau$, from now on). For that purpose, we used the routine \textit{modelador.x} included in the Stokes Inversion based on Response functions \citep[SIR,][]{RuizCobo1992} code.  Thus, all the plots in this work correspond to the atmospheric parameters at a constant optical depth (while the geometrical height may vary from pixel to pixel). Figure~\ref{FOV} shows the spatial distribution of the temperature for the two snapshots at a height where log~$\tau = 0$. We also highlight a smaller region with a coloured square in each panel which corresponds to the spatial domain we use for optimising the inversion configuration (orange) we use as a reference setup in the following sections or a region we study in detail in this work (blue).

\begin{figure*}
\begin{center} 
   \includegraphics[trim=0 0 0 0,width=18.2cm]{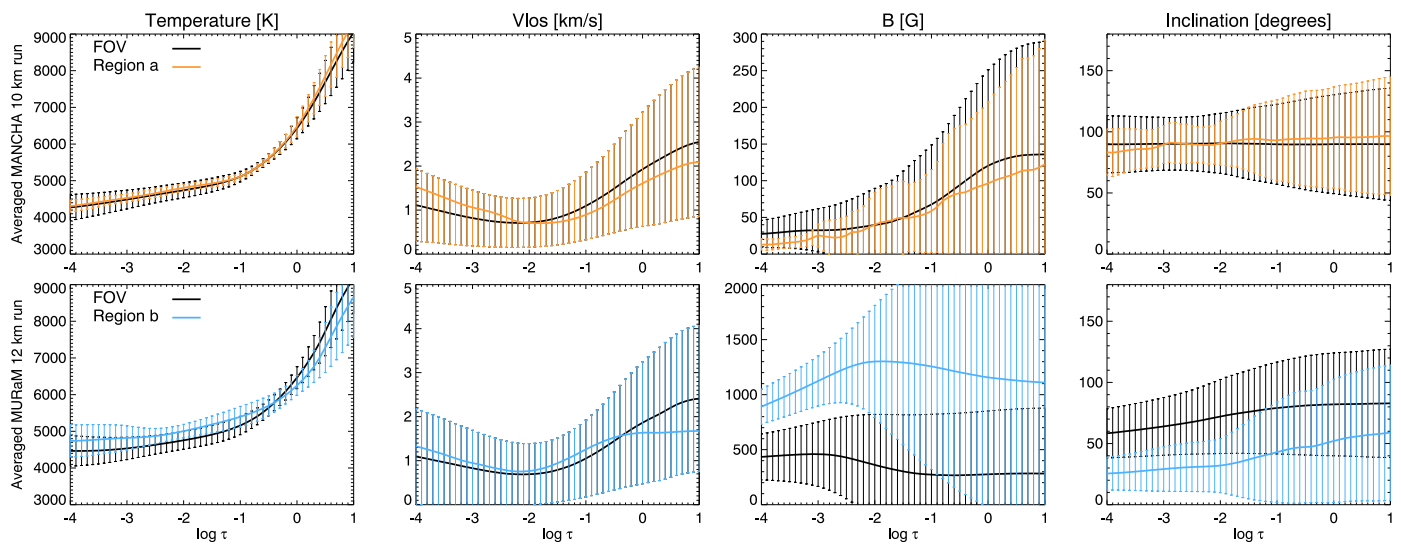}
 \vspace{-0.10cm}
 \caption{Evolution of the averaged atmospheric parameters with height. Top and bottom panels correspond to the MANCHA and MURaM runs presented in Figure~\ref{FOV}. Colours designate the averaged results over the small regions highlighted in the same figure. Bars display the standard deviation of the atmospheric parameters over the corresponding areas.}
 \label{Averaged}
 \end{center}
\end{figure*}

We mentioned above that this analysis aims to understand our capabilities for inferring atmospheric information from observations made by future generation 4m class telescopes like DKIST and EST. Both telescopes will have access to the 630~nm region in the visible, which is traditionally used by many telescopes. Therefore, in this first work, we focus only on this spectral region, including the \ion{Fe}{i} transitions at 6301.5 and 6302.5~\AA \, while additional spectral lines will be studied in the future. Furthermore, we use the atomic parameters presented in \cite{QuinteroNoda2021}, and we do not add any spatial degradation as the pixel size at 630~nm for a 4m telescope is comparable to that of the simulations used in this work:
\begin{dmath}
\textrm{pixel}_{\rm size} = \frac{\lambda}{2 \cdot D} \ [\textrm{rad}] = \frac{630\times10^{-9} \ [\textrm{m$\cdot$ rad}]}{2 \cdot 4.2 \ [\textrm{m}]}\times 206265\ \frac{[\textrm{arcsec}]}{[\textrm{rad}]} \times725 \ \frac{[\textrm{km}]}{[\textrm{arcsec}]}= 11.22 \ \textrm{km}
\end{dmath}

The synthesis of the Stokes profiles is done with SIR with a spectral sampling of 10~m\AA/pix over a spectral range that goes from -500 to 1500~m\AA \ from the rest wavelength of the line core of \ion{Fe}{i} 6301.5~\AA. The conditions mentioned above are easily achievable with the Visible Spectro-Polarimeter \citep{2022SoPh..297...22D} installed in DKIST, for example. We do not add any spectral degradation resulting from the finite spectral resolution of an instrument; for example, we do not convolve the spectrum by a point spread function of a given width. Furthermore, we do not consider additional instrumental effects like those produced by long integration times, limited slit width, or limited spectral sampling in the case of filtergraph instruments. However, a noise with different amplitude values is added to the Stokes parameters later on. The atomic abundance used during the process corresponds to that presented in \cite{Asplund2009}. In addition, although some authors have pointed out the necessity of including non-local thermodynamic equilibrium effects when modelling these spectral lines \citep[e.g., ][]{2020A&A...633A.157S}, we are going to make the synthesis and the inversion using the same forward module. Therefore, we do not expect to have significant issues in this regard. All the inversions are done using the Python wrapper presented in \cite{2021A&A...651A..31G}. We start the inversion from a modified version of the Harvard-Smithsonian reference atmosphere \citep[HSRA,][]{1971SoPh...18..347G}, which is designed to represent the inversion of actual observations as closely as possible; that is, the guess atmosphere we start the inversion from is independent of the simulation atmospheres used during the synthesis. The changes in the HSRA atmosphere include a constant LOS velocity and magnetic field vector so that SIR can apply multiplicative perturbations on those atmospheric parameters. The temperature-electron pressure relation is the original one published in the cited publication.

\subsection{General properties of the simulation runs}

The selected snapshots differ in many aspects, and so we believe it is beneficial to cover scenarios that could resemble strongly magnetised quiet Sun areas and weakly magnetised regions. We illustrate those differences in Figure~\ref{B_2D}, displaying the horizontal (B$\cdot \sin(\gamma)$, being $\gamma$ the inclination) and vertical (B$\cdot \cos(\gamma)$) components of the magnetic field at log~$\tau=0$. The quiet Sun run from the MANCHA code (left column) is characterised by weaker magnetic fields that populate most of the FOV with an intricate pattern. There are also areas with almost zero field strength that can represent a challenge for the inversion process. In the case of the MURaM Sunspot run (rightmost column), there are strongly magnetised areas with field strengths of over 2000~G. We believe they are an extension of the nearby sunspot. However, at the same time, those strong fields are surrounded by weak magnetic fields. Thus, the selected region resembles a plage region or small pore observation. We assume that fitting the Stokes profiles from those areas void of a magnetic field will also represent a challenge for the inversion process.

We complement the previous figure, adding the spatially averaged atmospheric parameters at every optical depth in Figure~\ref{Averaged}. We computed those values for the complete FOV of the MANCHA (top) and MURaM (bottom) runs and for the small regions highlighted with coloured squares in Figure~\ref{FOV}. In the case of the MANCHA run, both areas seem to be relatively similar with weak magnetic fields (B $<$ 100~G) and parallel to the solar surface ($\gamma\sim$ 90 degrees). Concerning the MURaM run, the averaged atmospheric parameters for the selected area and those for the entire FOV differ more significantly, with the atmospheric parameters of the enclosed region being more vertical with respect to the solar surface ($\gamma\sim$ 45 degrees). The enclosed region also shows a stronger magnetic field (B $>$ 1000~G). Still, in all the cases, we can see large deviations at each optical depth, indicating that the high spatial resolution of these runs produces significant small-scale variations (from one pixel to the neighbouring one), particularly in the magnetic field vector.

\subsection{Inversion configuration I: Nodes and inversion cycles}

One of the goals of this work is to estimate the loss of accuracy we suffer when inverting spectropolarimetric data with various noise levels. It is, therefore, paramount to find an inversion configuration that provides good results where the inferred atmosphere is as close as possible to the input one. For that purpose, as the two snapshots are complex, multiple inversion tests are required. However, each snapshot has more than 1000$\times$1000~pix$^{2}$, and so it is not feasible to perform all those inversion tests with the complete FOV within reasonable time limits. In that sense, we perform all the inversion tests described in this section, only inverting a small patch of the MANCHA run with a size of 100$\times$100~pix$^{2}$ (see the orange square in Figure~\ref{FOV}). In addition, although several authors have confirmed over the years \citep[among others,][]{2013ApJ...768...69B,2014A&A...569A..73Q,2021A&A...654A..11C} that using different initial atmospheres helps to improve the accuracy of the solution and reduce the chances of the code falling into a local $\chi^2$ minimum, in this work, we run every test with one atmosphere for simplicity and to reduce computational time. Therefore, we assume that the predictions we get here may be slightly improved using different guess atmospheres, which we plan to explore in more detail in future publications.

\begin{figure*}
\begin{center} 
   \includegraphics[trim=0 0 0 0,width=18.2cm]{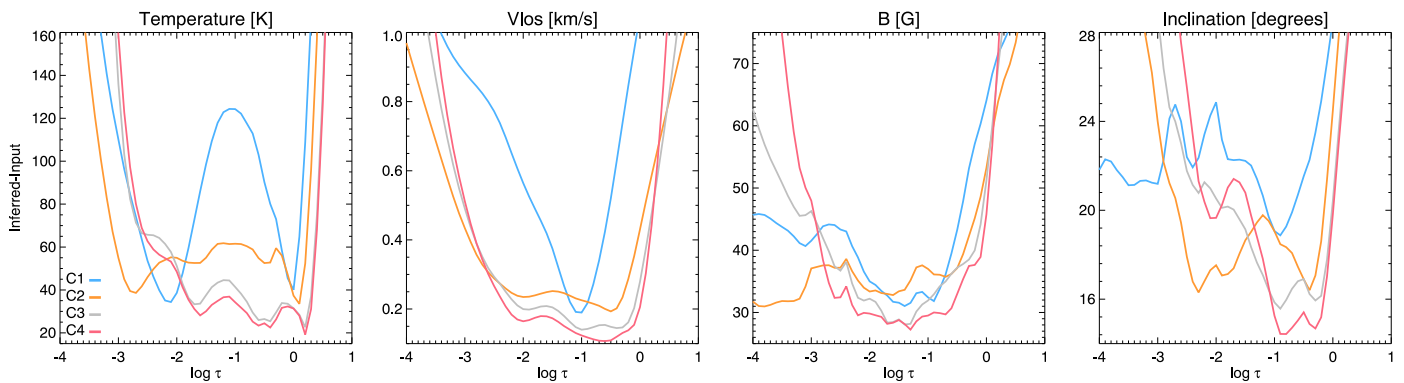}
 \vspace{-0.10cm}
 \caption{Average differences between the input and the inferred atmosphere at different optical depths. From left to right, we show the differences in temperature, LOS velocity, magnetic field strength, and inclination, respectively. Colours correspond to the various inversion configurations indicated in Table~\ref{Ctable}.}
 \label{Config}
 \end{center}
\end{figure*}

We start the study for the selection of nodes, that is, the heights where the guess atmosphere is perturbed and where the response functions (RFs) are evaluated using the values provided in Table~\ref{Ctable}. We follow a straightforward process to make the exercise as simple as possible and provide a reference guide for future observers interested in performing node-based inversions. We start from a simple configuration running one inversion cycle ---see C1--- changing the temperature at the top and bottom boundaries while the LOS velocity and magnetic field vector are constant with height. However, we do not aim to use these results as we know that the simulated atmospheres have physical parameters that strongly vary with height. Those variations with optical depth produce various asymmetries of the Stokes profiles \citep[among others, ][]{1997ApJ...482.1065B,1997ApJ...474..810M,2011A&A...530A..14V,2012ApJ...748...38S}. For this reason, we need more than 1 node on the LOS velocity and the magnetic field vector stratifications to fit the Stokes parameters with the highest accuracy. We therefore increase the number of nodes used for the inversion as we add more inversion cycles as described in Table~\ref{Ctable}. In each new iteration, we compare the input and the inferred atmosphere to determine whether or not we are improving the results by adding more nodes. We then compute the average differences, in absolute value, over the FOV used for these tests (the orange region in Fig.~\ref{FOV}) at each optical depth (the step is every  log~$\tau = 0.1$) to estimate the general accuracy of the fits. The results are shown in Figure~\ref{Config}.

\begin{center}
\begin{table}
\small
\begin{adjustbox}{width=0.40\textwidth}
  \bgroup
\def\arraystretch{1.25}
\begin{tabular}{|c|c|c|c|c|cccc}
        \hline 
 Parameter/Nodes   &  C1 &  C2 & C3 & C4     \\
        \hline
Temperature & 2 & 3 & 5 & 5 \\     
        \hline  
LOS velocity & 1 & 2 & 3 & 5 \\     
        \hline  
Field strength & 1 & 2 & 3 & 5 \\ 
        \hline  
Inclination & 1 & 2 & 3 & 5 \\  
        \hline  
Azimuth & 1 & 2 & 2 & 2 \\ 
        \hline  
Microturbulence & 1 & 1 & 1 & 1 \\ 
        \hline                                    
  \end{tabular}
  \egroup
\end{adjustbox}
\vspace{0.2cm}
\caption{Number of nodes used for the inversion of each atmospheric parameter. Each new column towards the rightmost side of the table adds (or maintains) more nodes for the inversion and corresponds to an additional inversion cycle; e.g. C1 is one cycle while C4 is done in four cycles, increasing the nodes as in C1-C3.} 
\label{Ctable}    
\end{table}
\end{center}

Starting with the results for the temperature, we can see that the first configuration, with constant perturbations along the LOS for the velocity and magnetic field, provides relatively good results at log~$\tau = 0$ and around log~$\tau = -2$ and much larger deviations between and outside those layers. We believe this is due to a lack of freedom, as we can see improvements in subsequent inversion cycles until we reach five nodes in C3 and C4. Interestingly, there is an improvement in C4, although the nodes for temperature are the same as in the previous inversion cycle. We believe it is because the rest of the atmospheric parameters have more freedom, and the fits are consequently better. In the case of the LOS velocity, we also see a continuous improvement from C1 to C4. For the field strength, we again see that the differences become smaller as we move from C1 to C4, with an average difference of less than 30 G for the latter configuration. Finally, for the magnetic field inclination, we see improvements again as we move from C1 to C4, with average differences of lower than 15 degrees in the latter case. 

After examining the previous results, we chose to continue our analysis using configuration C4 as it is better than the others in all the atmospheric parameters. However, we encourage the reader to examine the results in detail, particularly the fits of the Stokes profiles, when trying to define the inversion configuration. On the one hand, we need to add gradients to fit asymmetric profiles. On the other hand, deciding whether a linear gradient (2 nodes) is sufficient or more complex parabolic (3 nodes) or polynomial shapes (more than 3 nodes) are required is not entirely straightforward without an in-depth analysis of the results. Still, we recommend that users begin experimenting with the number of nodes presented here, as we find this to be a good baseline.

\begin{figure*}
\begin{center} 
   \includegraphics[trim=0 0 0 0,width=18.2cm]{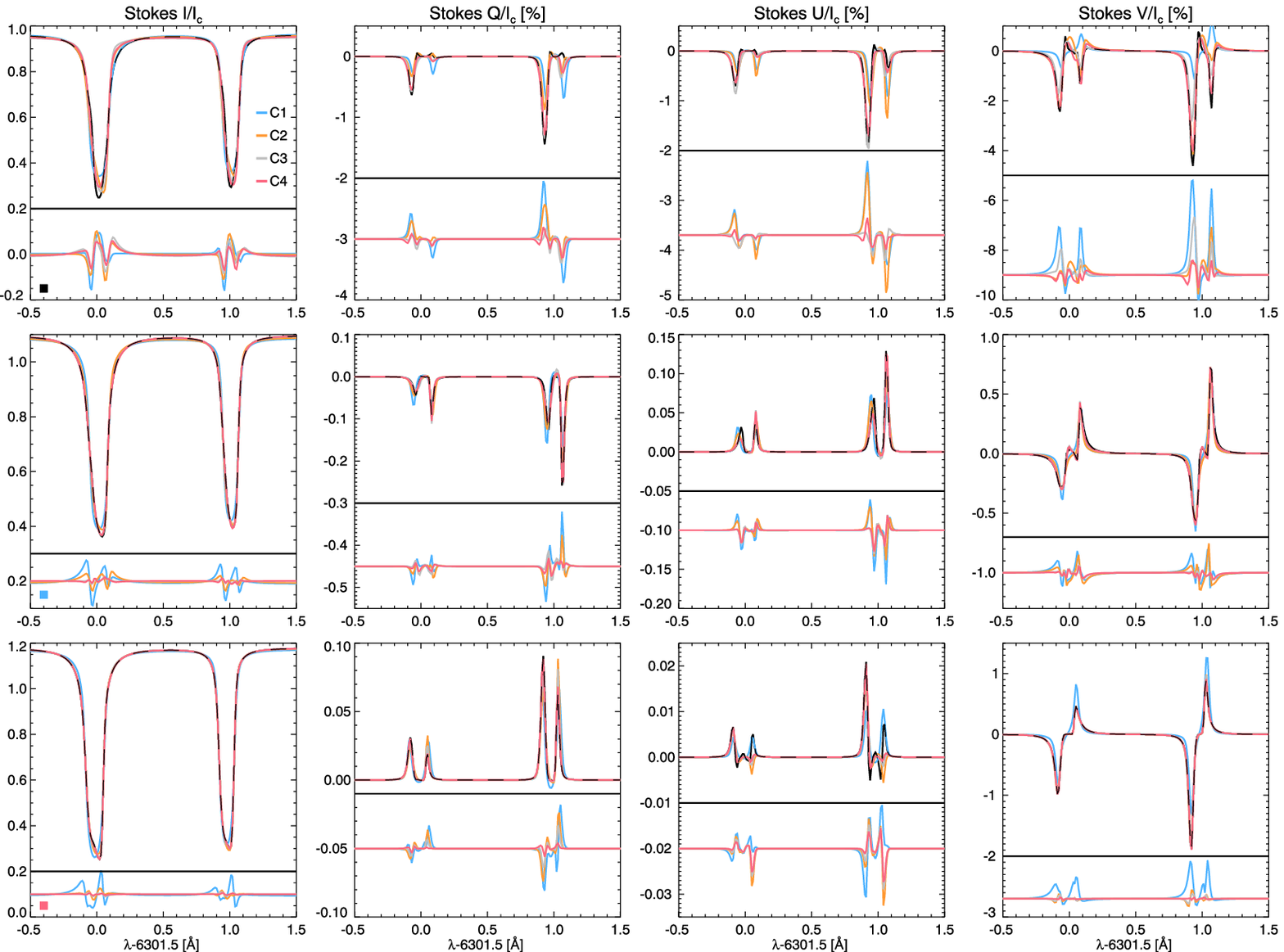}
 \vspace{-0.10cm}
 \caption{Stokes profiles corresponding to the three reference pixels (rows). From left to right, Stokes ($I$, $Q$, $U$, $V$), respectively. Black corresponds to the input profiles, while colours designate the Stokes parameters corresponding to the different node configurations presented in Table~\ref{Ctable}. Each panel includes the differences between the input and the inferred profiles below the horizontal solid black line. From top to bottom, each reference profile corresponds to an intergranular lane, edge of a granule, and top of a granule, respectively. We show the exact location in Figure~\ref{Mancha2D_a}, where we provide a zoom onto the region highlighted in the top panel of Figure~\ref{FOV}.}
 \label{Profiles}
 \end{center}
\end{figure*}

\begin{figure*}
\begin{center} 
   \includegraphics[trim=0 0 0 0,width=18.2cm]{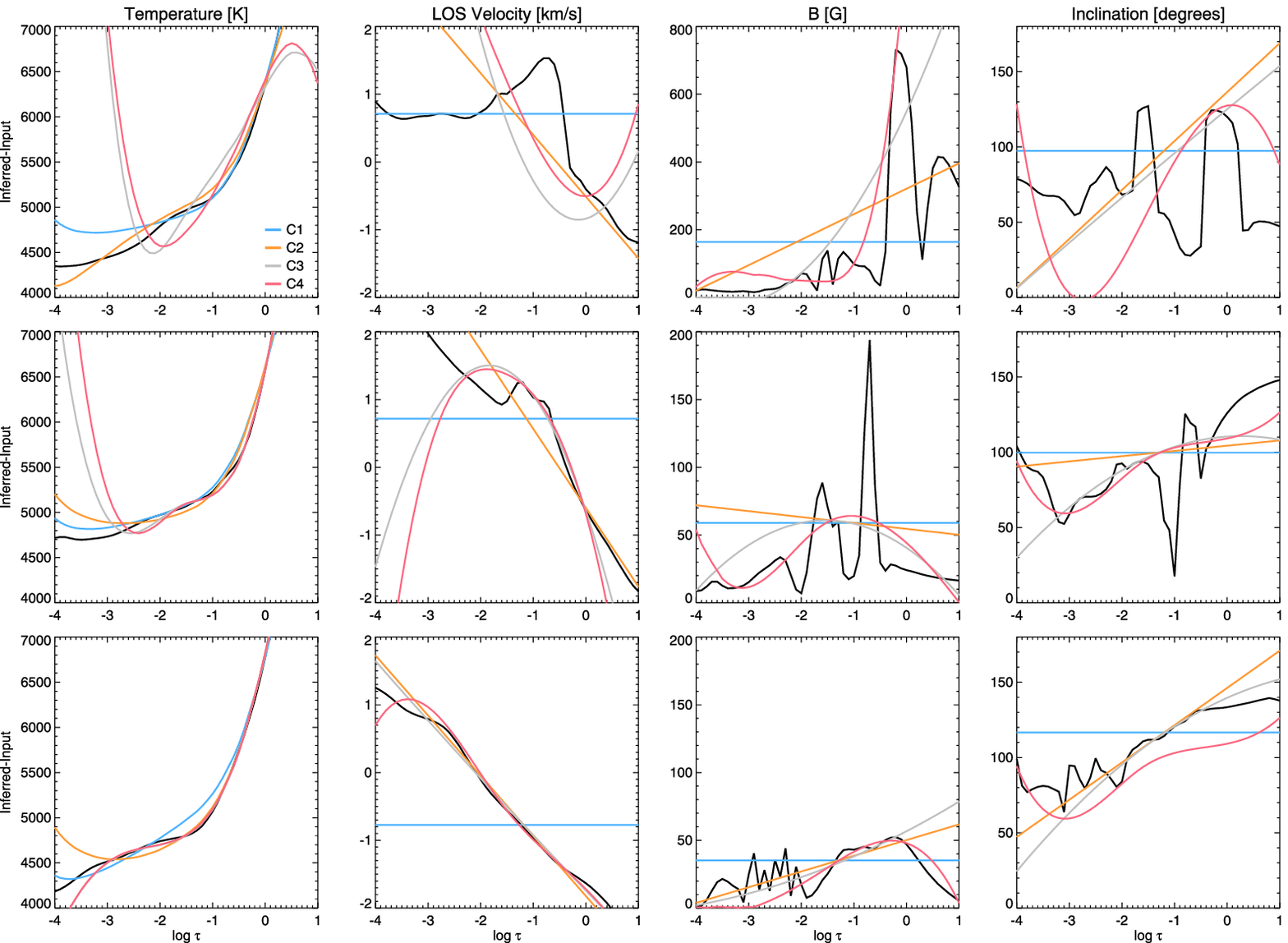}
 \vspace{-0.10cm}
 \caption{Atmospheres corresponding to the pixels presented in Figure~\ref{Profiles} (different rows). We show the temperature, LOS velocity, magnetic field strength, and inclination from left to right. Black corresponds to the input profiles, while colours designate the Stokes parameters corresponding to the different node configurations presented in Table~\ref{Ctable}.}
 \label{Atmospheres_1D}
 \end{center}
\end{figure*}

Finally, we want to mention that we do not pay particular attention to the magnetic field azimuth in this work. This is mainly because the expected magnetic fields are weak in strength and are strongly local (maybe occupying a few pixels in size), and so it would be challenging to solve the 180 degree ambiguity in order to interpret the inferred azimuth values reliably \citep[e.g., ][]{2006SoPh..237..267M}. Instead, could check the input atmosphere and, based on that, add or subtract 180 degrees to match the correct orientation of the magnetic field. Also, we could compare the cosine of the azimuth (that will range from 0 to 1), removing the effect of the 180 ambiguity. However, we prefer to leave the azimuth comparison out of this work, and we will address this subject ---and the different  options to reliably infer it in this complex numerical scenario--- in future publications. Still, we want to emphasise that the atmospheric parameters are fitted using the node values presented in Table~\ref{Ctable} to get good fits for the Stokes $Q$ and $U$ profiles, and hence the most accurate atmosphere for all the physical quantities.

\subsection{Inversion configuration II: Profiles and atmospheres}

We mentioned in the previous section that examining the fits of a few reference profiles is advisable in order to gain confidence when choosing the inversion configuration that works best for our goals. In that sense, we show in Figure~\ref{Profiles} the fit for three reference pixels where the code fits the profiles relatively poorly (top), relatively well (middle), and with high accuracy. Furthermore, we show the differences between the input (black) and the inferred Stokes parameter (colours) at the bottom of each panel. We again see significant differences for C1 (blue) and C2 (orange) and much lower deviations for C3 (grey) and C4 (red). In general, the latter two configurations behave very well. Still, in a few cases, C4 produces smaller differences, and in the rest of the cases, the results are similar, indicating that C4 provides the most accurate results for our scenario. Additionally, we can investigate further, comparing the input atmosphere that produced those profiles with that obtained by the code when using each inversion configuration.

\begin{figure}
\begin{center} 
   \includegraphics[trim=0 0 0 0,width=8.5cm]{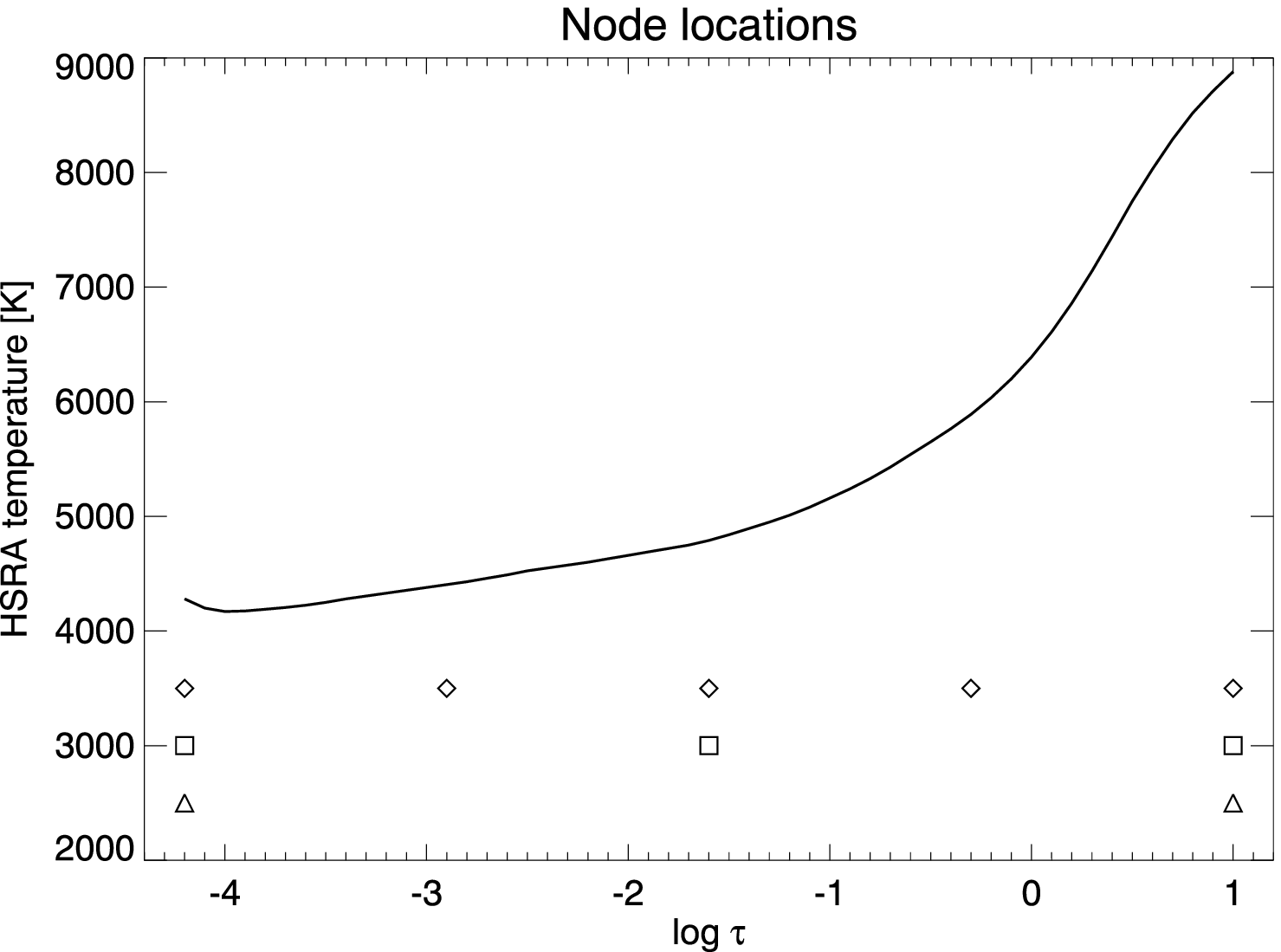}
 \vspace{-0.10cm}
 \caption{Reference location of nodes in SIR. The cases corresponding to 2, 3, and 5 nodes (see Table~\ref{Ctable}) have nodes located as shown by triangles, squares and diamonds, respectively. As a reference, we include the temperature of the HSRA atmosphere.}
 \label{Nodes_position}
 \end{center}
\end{figure}

In Figure~\ref{Atmospheres_1D}, we present the comparison for temperature, LOS velocity, and magnetic field strength and inclination for the three reference pixels shown in Figure~\ref{Profiles}. Starting with the temperature, we have relatively smooth stratifications that are well recovered by all the configurations. In this case, we cannot see significant differences between C3 and C4. This is probably because the number of nodes is the same in both cases. Something similar happens for the LOS velocity, with both configurations showing approximately the same behaviour and more accurate results than C1 and C2. Finally, we see slightly better accuracy for C4 in the magnetic field strength and inclination, although there are no significant differences. Therefore, based on these results, combined with the previous analysis, we continue using C4. However, we would not object to another user preferring to use C3 simply because the number of nodes is lower, considering that the fit of the Stokes profiles (the only measure of accuracy when working with observations) is slightly poorer than with C4.

Finally, before examining the inversion configuration, we need to comment on the general accuracy of the fits and the atmospheres we are inferring. It is clear that in some cases (Figure~\ref{Atmospheres_1D}), the code struggles to reproduce the complex atmosphere from the 3D simulation. This lack of accuracy is something we expected in the case of having atmospheres that change over short height scales. The code uses several nodes (up to five in our case) that are equidistant in optical depth (see Figure~\ref{Nodes_position} as a reference), and so it is not possible to produce such complex stratifications. 

\begin{figure*}
\begin{center} 
   \includegraphics[trim=0 0 0 0,width=18.2cm]{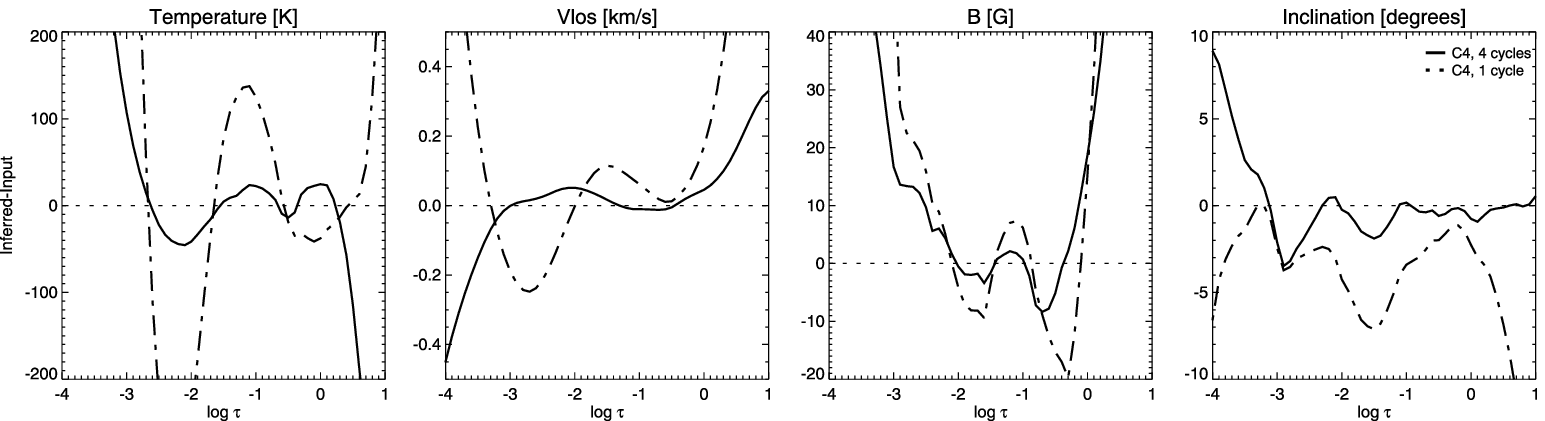}
 \vspace{-0.10cm}
 \caption{Average differences between the input and the inferred atmosphere at different optical depths. From left to right, we show the differences for temperature, LOS velocity, magnetic field strength and inclination, respectively. Line styles correspond to the case where we reach configuration C4 in Table~\ref{Ctable} after running the previous cycles (solid) or to cases where we run a single inversion cycle using the nodes described in column C4 of Table~\ref{Ctable} starting from the modified HSRA atmosphere (dashed).}
 \label{Cycles}
 \end{center}
\end{figure*}

At the same time, it is not possible from a radiative transfer point of view to have sensitivity to such small-scale variations with our observation configuration. Spectral lines are sensitive to a given range of optical depth layers that it usually not as narrow as the changes shown by the simulation. In this case, we only have two spectral lines that are sensitive within a similar range of heights \citep[see, e.g., ][]{QuinteroNoda2021}. Therefore, it is impossible to reach the degree of complexity shown by the simulation, although the code generally does relatively well. 

\subsection{Inversion configuration III: Iteration cycles}

At this point, it may not be clear as to whether or not we need to run multiple inversion cycles, increasing the number of nodes each time, for example, as we move from C1 to C4 in Table~\ref{Ctable}, or if we could run one inversion cycle using the configuration C4. Therefore, we compare both scenarios in Figure~\ref{Cycles}. 

The results for the case where we run four inversion cycles, gradually increasing the number of nodes in each cycle, are presented in solid black. At the same time, the dashed line shows the result with one inversion cycle using the nodes specified by C4 and starting from the modified HSRA atmosphere. We start in both cases from a modified HSRA atmosphere where LOS velocity and magnetic field vector are constant with height. In the first example, the code changes the temperature at the top and bottom boundary in C1 and applies a constant perturbation in the rest of the atmospheric parameters. In the next cycle, the code improves the previous atmosphere, changing its temperature at the bottom, the top and the middle of the atmosphere, and at the bottom and the top for the rest of the parameters (adding gradients along the LOS for the velocity and the magnetic field vector). From there, in each inversion cycle, the code starts from a perturbed atmosphere, which delivered reasonable fits, and changes it again based on the RF of the previous atmosphere. Each iteration gradually adds more freedom in incremental steps. In the second study, that is, starting with C4 directly and running one inversion cycle, the code probably has too much freedom when starting from an atmosphere (modified HSRA) that is far from the target solution; this means the inversion iterations that the code can make will therefore allow a solution to be achieved that, in general, is poorer than in the case of gradually increasing the freedom of the atmospheric parameters. Therefore, we encourage the users of node-based inversion codes to gradually increase the number of nodes in a manner similar to that presented in Table~\ref{Ctable}.

\begin{figure*}
\begin{center} 
   \includegraphics[trim=0 0 0 0,width=18.2cm]{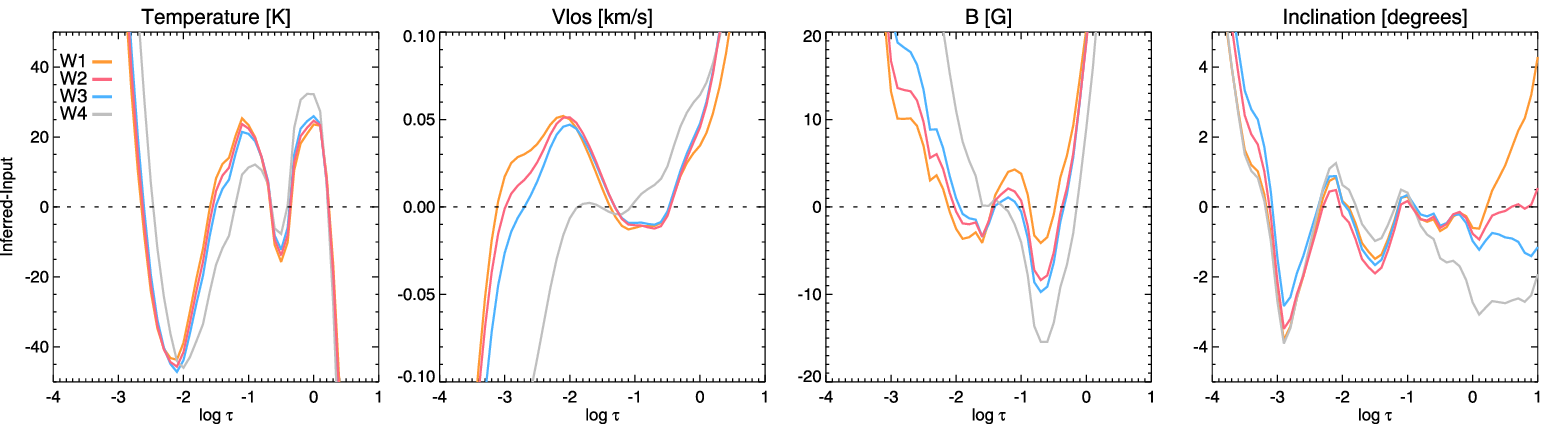}
 \vspace{-0.10cm}
 \caption{Average differences between the input and the inferred atmosphere at different optical depths. From left to right, the differences for temperature, LOS velocity, magnetic field strength, and inclination, respectively. Colours correspond to the inversion configurations indicated in Table~\ref{Wtable}.}
 \label{Weights}
 \end{center}
\end{figure*}

\subsection{Inversion configuration IV: Weights}

We evaluate whether or not we can improve the previous results using a different configuration in this section. SIR is an inversion code that synthesises the Stokes profiles from different atmospheric models until the synthetic spectrum matches the observed one \citep[see, for instance,][]{2022A&A...660A..37R}. The process implies solving the radiative transfer equation multiple times while perturbing the atmospheric parameters until an accurate solution is achieved, that is, one where the difference between synthetic and observed profiles is as small as possible for the entire analysed spectral range. That accuracy is defined by the following $\chi^2$ merit function:
\begin{equation}
\chi^2 = \frac{1}{N_f}\sum_{S=1}^{4}\sum_{i=1}^{n_{\lambda}}\left[ I_s^{\rm Obs}(\lambda_i)-I_s^{\rm Syn}(\lambda_i)       \right]^2 w_s(\lambda_i)^2
,\end{equation}
where index $S$ runs over the four Stokes parameters, $i$ covers the wavelength samples up to $n_\lambda$, and $N_f$ stands for the number of degrees of freedom, that is, the difference between the number of observables (four Stokes profiles $\times$ $n_\lambda$) and the number of free parameters used in the inversion (the sum of all the values included in each column in Table~\ref{Ctable}). The weights $w_s(\lambda_i)$ are `best-fit normalisation' factors used to add more emphasis to a specific Stokes profile, for example, Stokes~$V$, or a specific spectral line, when fitting multiple spectral lines simultaneously, promoting better fits of those individual features. In this section, knowing that the target is to fit weak polarisation signals in a quiet Sun scenario, we study the correlation between the input and the inferred atmosphere when we add more weight to the Stokes $Q$ and $U$ profiles.

\begin{figure*}
\begin{center} 
   \includegraphics[trim=0 0 0 0,width=18.2cm]{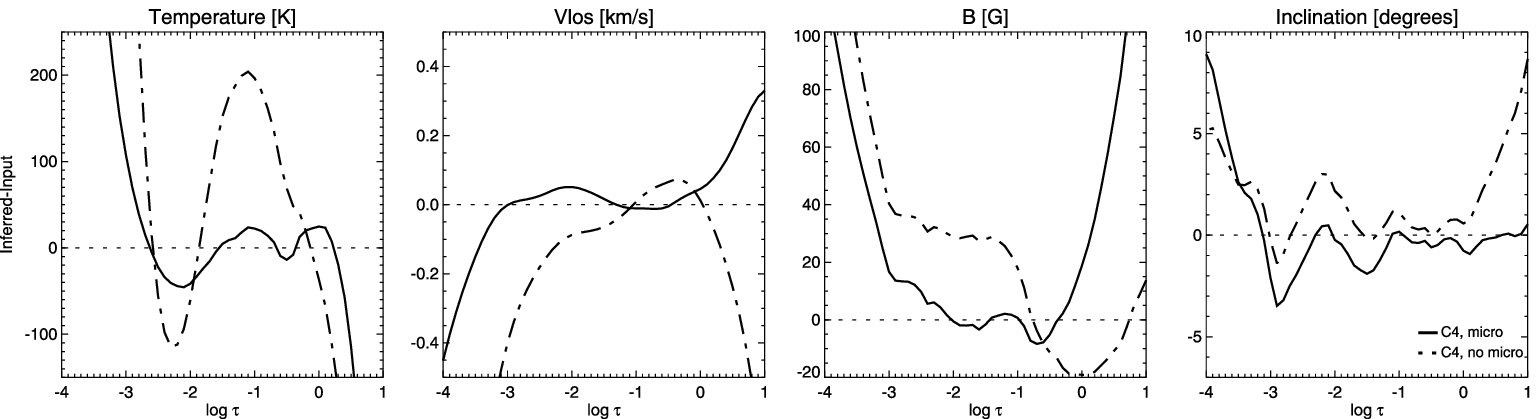}
 \vspace{-0.10cm}
 \caption{Average differences between the input and the inferred atmosphere at different optical depths. From left to right, the differences for temperature, LOS velocity, magnetic field strength, and inclination, respectively. Line styles correspond to the configuration C4 in Table~\ref{Ctable} and W2 in Figure~\ref{Weights} when setting the microturbulence as a free parameter (solid) or not (dashed).}
 \label{Micro}
 \end{center}
\end{figure*}

We compare the four different weight configurations presented in Table~\ref{Wtable}. In each case, we run the inversion using configuration C4, increasing the weight given to the Stokes $Q$ and $U$ parameters. The results are presented in Figure~\ref{Weights}. We can see that in the case of temperature, the differences are similar except for W4 (where the weights are much higher for the linear polarisation parameters), which shows greater differences in general. This higher difference means that the weight is so significant for Stokes $Q$ and $U$ that the code does not fit Stokes $I$ with reasonable accuracy. Hence, the greater deviations in the temperature. We do not recommend reaching this point, even if we are primarily interested in the magnetic field vector, because, in the end, the Stokes $Q$, $U$, and $V$ profiles are derivatives of the Stokes $I$ parameter. In other words, a poor fit for Stokes $I$ will also mean poor results for the rest of the Stokes parameters. We can see this trend for the rest of the atmospheric parameters, where the grey line produces a lower correlation for the magnetic field strength and LOS velocity. It is comparable to the other solutions for the inclination, which means that the code accurately fits Stokes $Q$ and $U$. Still, we are losing accuracy in the rest of the atmospheric parameters. It is therefore essential to bear in mind that weights should be chosen carefully, in this case, emphasising Stokes $Q$ and $U$ because the magnetic field is weak and inclined (see Figure~\ref{B_2D}) but without noticeable degradation of the fit of Stokes $I$ because the error will extend to the inferred temperature, velocity, and field strength. Therefore, the most balanced solutions are obtained using W2 and W3. Of the two, W2 (red) is slightly better for the magnetic field strength, but W2 and W3 are similar for the inclination and the temperature. We therefore plan to use W2 combined with C4, although we recommend users experiment with weights until the inversions provide good fits on average for the pixels of interest.

\begin{center}
\begin{table}
\small
\begin{adjustbox}{width=0.40\textwidth}
  \bgroup
\def\arraystretch{1.25}
\begin{tabular}{|c|c|c|c|c|cccc}
        \hline 
 Parameter/Weights   &  W1 &  W2 & W3 & W4     \\
        \hline
Stokes $I$ & 2 & 2 & 2 & 2 \\     
        \hline  
Stokes $Q$ & 1 & 10 & 20 & 100 \\     
        \hline  
Stokes $U$ & 1 & 10 & 20 & 100 \\ 
        \hline  
Stokes $V$ & 5 & 5 & 5 & 5 \\  
        \hline                               
  \end{tabular}
  \egroup
\end{adjustbox}
\vspace{0.2cm}
\caption{Weight options used for the inversion of each atmospheric parameter. The inversion configuration is fixed and corresponds to C4 in Table~\ref{Ctable}.} 
\label{Wtable}    
\end{table}
\end{center}

\subsection{Inversion configuration V: Microturbulence}

The microturbulence is a correction added when computing the Doppler broadening; in SIR, it is defined in the Doppler width, $\Delta \lambda_D$ in the following way:
\begin{equation}
\Delta \lambda_D=\lambda/c  \sqrt{\xi^2+\frac{2kT}{M}},
\end{equation}
where $\lambda$ is the spectral line rest wavelength, $\xi$ is the microturbulence, $T$ the temperature, $M$ the mass of the atom involved in the transition, and $c$ and $k$ are the speed of light and the Boltzmann constant, respectively.

There is no complete consensus as to why the microturbulent velocity is used when fitting observations \cite[see, for instance, Sec.~3 in][]{2019A&A...630A.133M}. However, we still add it as a free parameter in inversions to account for the complicated vertical stratification in the LOS velocity, which cannot be fully resolved by node-based inversion codes, mainly in the present scenario where the numerical simulations were computed with 7 and 8~km vertical resolution, respectively. Therefore, in this section, we evaluate the accuracy of the fits when using or not microturbulence as a free parameter. We reiterate here that we did not incorporate any microturbulence correction during the synthesis process (the original simulation runs do not include one). However, similarly to what was shown in \cite{2019A&A...630A.133M} and \cite{2022A&A...660A..37R}, we expect the results to be more accurate when using it as a free parameter.

Figure~\ref{Micro} shows the average differences between the input and the inferred atmosphere when inverting the microturbulence with a constant value with height (solid) or not (dashed). Results for temperature are more accurate in the whole domain, indicating that the fits of Stokes $I$ are also more accurate. This improved accuracy also translates to better fits in general for the rest of the Stokes profiles and smaller differences for all the atmospheric parameters. Therefore, as previous works have found, the inversion code provides better results when using microturbulence as a free parameter. We believe the main reasons for this behaviour are either that the degrees of freedom in the temperature stratification are insufficient in number (which is unlikely for this case) or that there is a strong coupling or degeneracy between temperature and microturbulence. In any case, and given our commitment to exploring these results in future works, we continue using configuration C4 with one node for the microturbulence in this publication.

\begin{figure*}
\begin{center} 
   \includegraphics[trim=0 0 0 0,width=17.5cm]{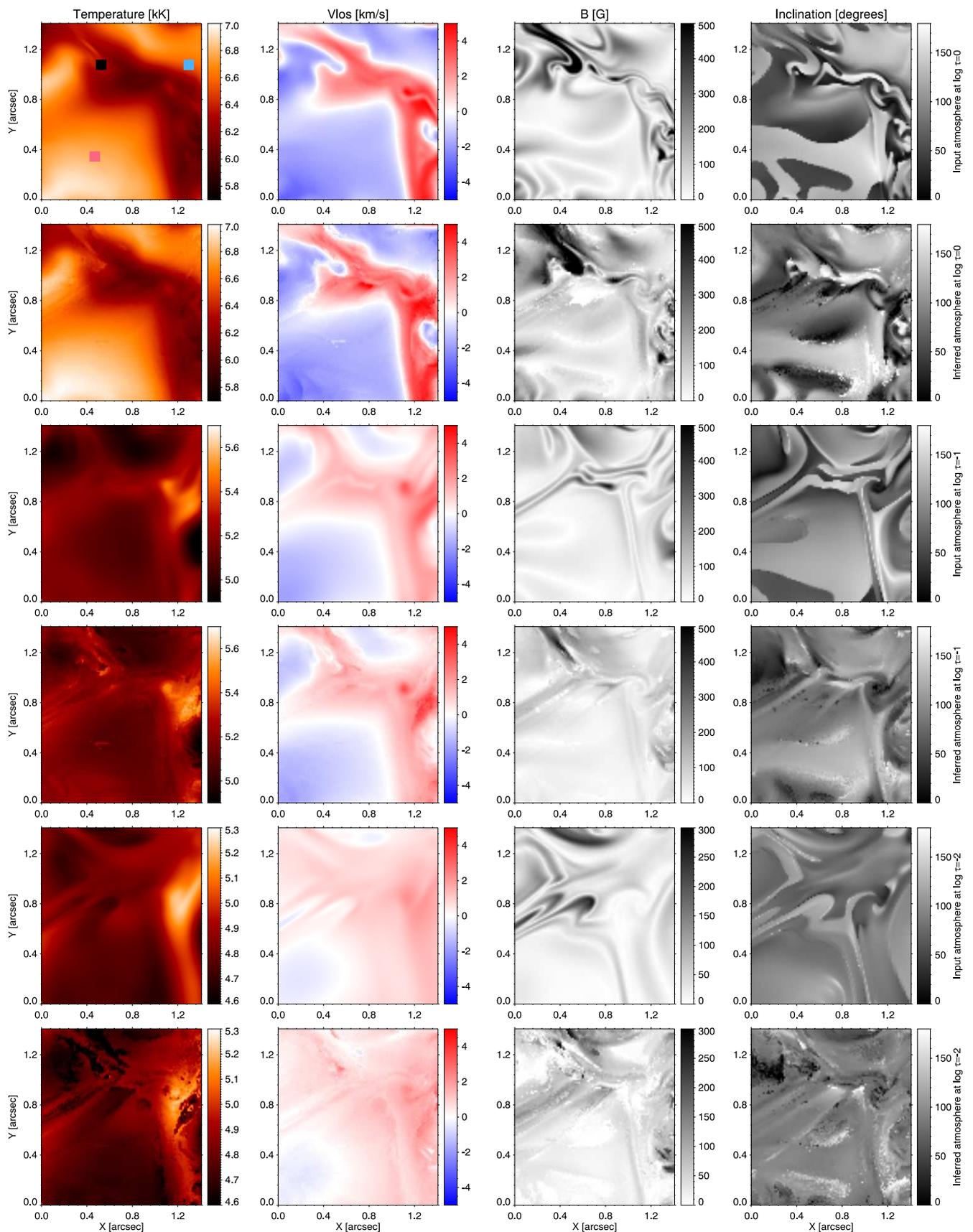}
 \vspace{-0.10cm}
 \caption{Comparison between the input atmosphere (odd rows) and the one inferred with SIR using configuration C4 and the weights W2 (even rows). From left to right, we show the temperature, LOS velocity, magnetic field strength, and inclination. From top to bottom, we display the spatial distribution of the atmospheric parameters at three reference layers at log~$\tau_{500}$=[0,-1,-2]. The FOV corresponds to the spatial domain within the square region (orange) in the top panel of Fig.~\ref{FOV}. Coloured squares in the top leftmost panel designate the locations of the pixels studied in Figures~\ref{Profiles} and \ref{Atmospheres_1D}. Black, blue, and red correspond to the top, middle, and bottom rows in the cited figures, respectively.}
 \label{Mancha2D_a}
 \end{center}
\end{figure*}

\begin{figure*}
\begin{center} 
   \includegraphics[trim=0 0 0 0,width=18.2cm]{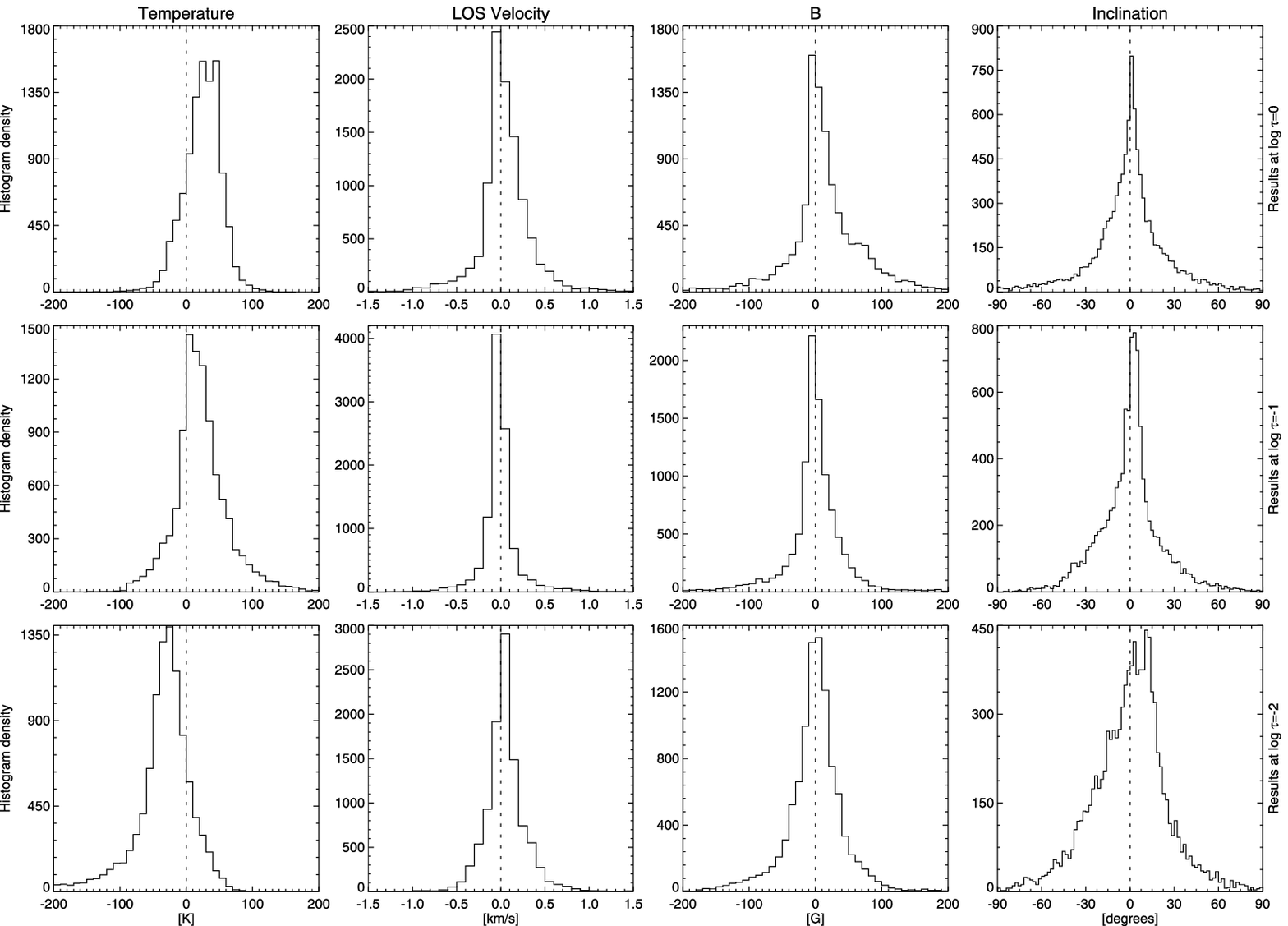}
 \vspace{-0.10cm}
 \caption{Histogram containing the differences between the inferred and the input atmosphere at selected optical depths for the region enclosed by the orange square in Figure~\ref{FOV}. The differences in temperature, LOS velocity, magnetic field strength, and inclination are shown from left to right. Each row corresponds, from top to bottom, to reference layers at log~$\tau_{500}$=[0,-1,-2]. The bin size is 10~K, 0.1 km/s, 10~G, and 2~degrees, respectively.}
 \label{Mancha_a_histo}
 \end{center}
\end{figure*}

\subsection{Inversion configuration VI: Spatial distribution of the atmospheric parameters}

Above, we focused on the average differences between the input and the inferred atmospheres to define the inversion configuration. In this section, we examine the accuracy of the results in more detail, studying the spatial distribution of the atmospheric parameters at different optical depth layers. We use three atmospheric layers at log~$\tau$ = [0, -1, -2] as a reference, because they cover the typical range of maximum sensitivity for the spectral lines at 630~nm \citep[e.g.,][]{QuinteroNoda2021}. Finally, we compare the input and the inferred atmospheres using the node configuration C4 and the weights W2 in Figure~\ref{Mancha2D_a}.

Starting with the temperature, we can see that the results at log~$\tau=0$ are very accurate, with almost no apparent differences between the input and the inferred atmosphere. We see a similar behaviour for higher layers, that is, log~$\tau=-1$, although some pixels deviate from the input atmosphere. This deviation continues increasing at log~$\tau=-2$, where some areas appear with much lower temperatures in the inferred atmosphere (see around [0.4,1.2] arcsec). 

In the case of the LOS velocity, we see that the inversion results closely resemble that of the input atmosphere in all layers, even at log~$\tau=-2$, where only minor differences can be found. 

Magnetic field strength results at log~$\tau=0$ are reasonably accurate with many of the small-scale features recovered in the inversion, although there are more significant differences at around  [0.6, 0.9] arcsec. This behaviour is also found at log~$\tau=-1$, with the inversion reproducing the complex and thin structures found in the selected FOV but also providing stronger magnetic fields in areas where the input parameter was weak (see ---e.g.--- [0.4, 1.2] arcsec). Higher up, at log~$\tau=-2$, the results resemble the input atmosphere, but the general accuracy has dropped substantially. Finally, the magnetic field inclination is well recovered at log~$\tau=0$, with the inferred atmosphere following the intricate patterns where the magnetic field changes rapidly, even from one polarity to the opposite one, on spatial scales of a few pixels. The same happens for higher layers, with a general landscape that resembles the complexity of the input atmosphere, although that substantially drops in accuracy at log~$\tau=-2$.

\subsection{Inversion configuration VII: Accuracy of the inferred atmosphere}

In order to investigate the accuracy of the fits in more detail, in Figure~\ref{Mancha_a_histo} we show a histogram containing the differences for all the pixels in the selected FOV between the input and the inferred atmosphere at the same optical depth layers examined in the previous figure. Again, we are interested in the accuracy for inferring all the atmospheric parameters, and so we show, from left to right, the results for temperature, LOS velocity, magnetic field strength, and inclination. 

Starting with temperature, we obtain results that peak close to zero but show that the inversion process tends to obtain slightly hotter atmospheres than the input ones. However, this trend is less pronounced at log~$\tau=-1$, where the peak is much closer to zero. Furthermore, this trend is reverted at log~$\tau=-2$ where the code provides slightly cooler atmospheres. In any case, the width of the histogram is relatively narrow, having a Gaussian-like shape with wings that drop substantially before the $\pm100$~K mark. As mentioned in the previous section, LOS velocity results are very close to the input atmosphere at most layers. This can be seen with a narrow Gaussian-like function for the differences that peaks close to zero in the three selected atmospheric layers. In the case of the field strength, differences also peak at zero, although we see an extended right tail at log~$\tau=0$, indicating that multiple pixels show larger amplitude for the field strength than the original ones. However, this tail is less prominent in the upper layers, indicating that the inferred magnetic field strength is close to the original one in general. Finally, for the inclination, we again see that the differences display a Gaussian-like profile that peaks at zero, although it has extended wings, mainly at log~$\tau=-2$. In those cases, the differences could be larger than 30 degrees, which can be considered as low accuracy for the magnetic field inclination. Still, in general, mainly at lower layers, we have a prominent peak at zero, pointing out that the fits are generally accurate.

\begin{figure*}
\begin{center} 
   \includegraphics[trim=0 0 0 0,width=18.2cm]{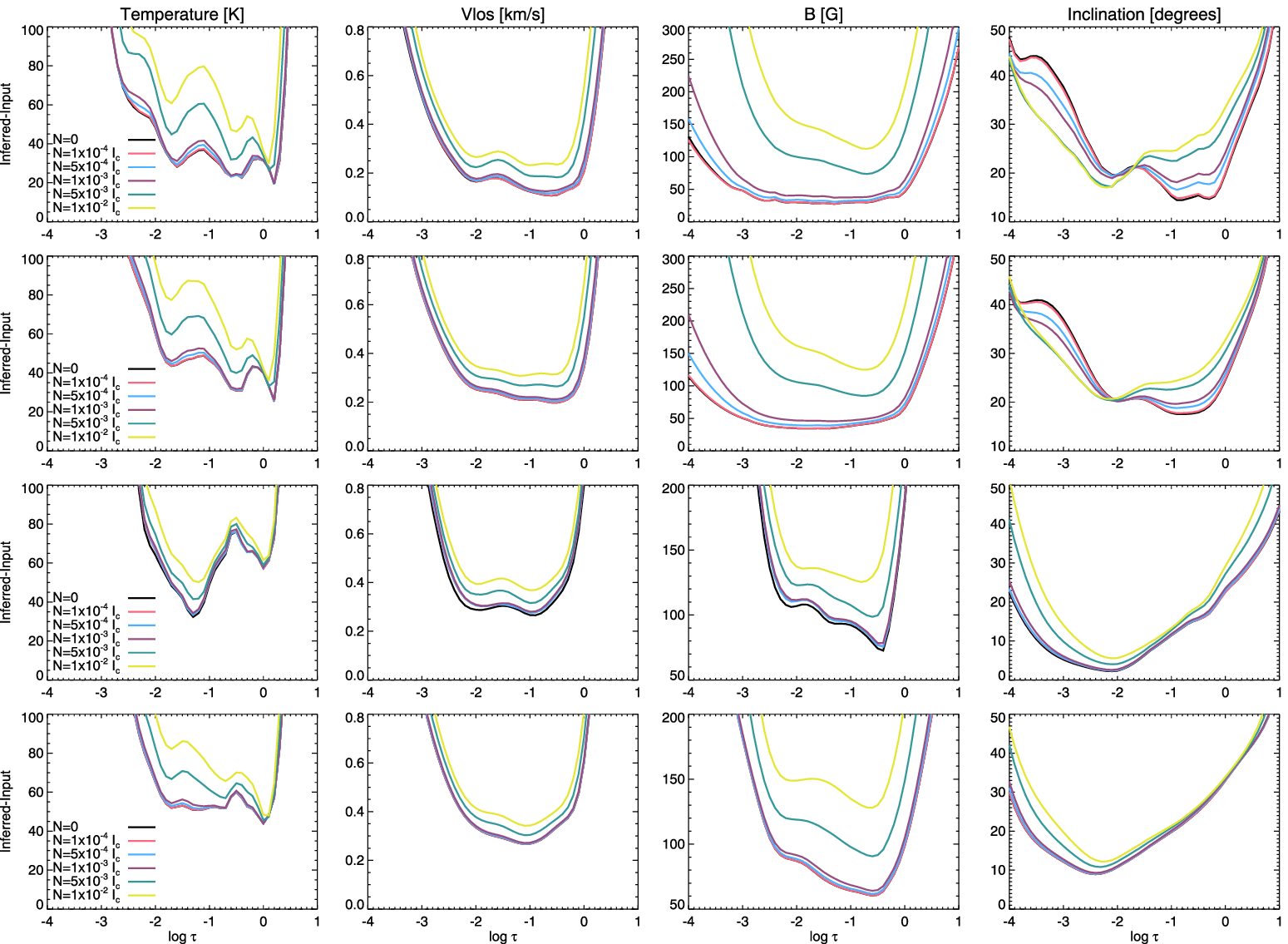}
 \vspace{-0.10cm}
 \caption{Average differences between the input and the inferred atmosphere at different optical depths when adding noise to the synthetic profiles. From left to right, we show the differences for temperature, LOS velocity, magnetic field strength, and inclination, respectively. Colours correspond to the noise values indicated in the leftmost column. Rows are associated with different reference FOVs, with the top, second, third, and the bottom row corresponding to the region inside the orange box, the complete Mancha FOV, the region inside the blue box, and the complete FOV of the MURaM run presented in Figure~\ref{FOV}, respectively.}
 \label{Noise_comparison_total}
 \end{center}
\end{figure*}

\section{Results for noisy spectra}

The strategy for the following sections is similar for each selected FOV and simulation run. First, we add a noise degradation, which is different for every wavelength and Stokes parameter. Next, we modify the amplitude of the noise using the following values: [0, 1$\times10^{-4}$, 5$\times10^{-4}$, 1$\times10^{-3}$, 5$\times10^{-3}$, 1$\times10^{-2}$] of $I_c$. We then invert the Stokes profiles using the configuration C4 and the weights W2 (see Tables~\ref{Ctable} and \ref{Wtable}) for each noise amplitude and compute the average differences over the whole spatial domain at each optical depth. We start with the highlighted area (see the orange box in the leftmost column of Figure~\ref{B_2D}) and the complete FOV of the Mancha run, and then move to the selected area (see the blue box in the rightmost column of Figure~\ref{B_2D}), and the complete FOV of the MURaM run. The results for the inversion of the weakly magnetised areas in the Mancha run, that is, the same region used in Section~\ref{Method}, are displayed in the first row of Figure~\ref{Noise_comparison_total}. Each coloured line represents a different noise level, from no noise in black to 1$\times10^{-2}$ of $I_c$ in yellow.

Starting with the temperature, we can see that the differences are similar to low noise values and begin to increase as we increase the noise, being noticeably higher at 5$\times10^{-3}$ of $I_c$ and with larger noise values. Interestingly, the same pattern is found for the rest of the atmospheric parameters with similar and small differences for noise levels up to 1$\times10^{-3}$ of $I_c$. Only the inclination shows more variations for different noise values, where only a noise level of 1$\times10^{-4}$ of $I_c$ (red) is very close to the case without noise (black). The same pattern is found when inverting the Stokes profiles over the entire FOV presented in the top row of Figure~\ref{FOV}, with similar differences (second row in Figure~\ref{Noise_comparison_total}) for most of the atmospheric parameters when the noise levels are below 1$\times10^{-3}$ of $I_c$. Only the inclination, once again, is more sensitive to the noise values and shows noticeable deviations when the noise is larger than $\times10^{-4}$ of $I_c$.

The results for the inversion of the strongly magnetised area in the MURaM run are displayed in the third row of  Figure~\ref{Noise_comparison_total}. In this case, the results are similar to the previous case, that is, differences are mainly noticeable when the noise is larger 1$\times10^{-3}$ of $I_c$. However, in this scenario, the same pattern can be seen for the inclination with more significant deviations only when reaching 5$\times10^{-3}$ of $I_c$ and 1$\times10^{-2}$ of $I_c$. This applies to the case where we only choose a strongly magnetised area (see third row) like the one highlighted in blue in the rightmost column of Figure~\ref{B_2D}, and to the case where we use the entire FOV. This points out that when the average field strength is relatively large on average (see the bottom row in Figure~\ref{Averaged}), the accuracy of the results ---even for the inclination--- is still high when the noise levels are on the order of 1$\times10^{-3}$ of $I_c$. Finally, it is worth mentioning that the black (noiseless) and red (1$\times10^{-4}$ of $I_c$) lines are almost identical in all the examined scenarios, indicating that the maximum accuracy the code can provide for the present configuration (e.g. node selection, spectral lines, etc.) is reached with noise values of that order.

\section{Summary}

In this work, we study the accuracy with which we can infer atmospheric information from realistic numerical MHD simulations that reproduce the spatial resolution we can achieve with future observations using the 4m class telescopes DKIST and EST. We used two intrinsically different simulation runs from different numerical codes (MANCHA and MURaM), simulating different scenarios: a weakly magnetised area and a sunspot. The aim is to increase the applicability of the analysis. We studied multiple inversion configurations until we obtained a good correlation between the input and the inferred atmosphere in a wide range of heights. We hope the thorough explanations provided in this step of the process will help future users of inversion codes define an inversion configuration that matches their needs and gets the most from their observations. Also, to continue providing a reference for future observing proposals, we examine how the inversion accuracy depends on the noise level of the Stokes profiles. The results indicate that when the majority of the inverted pixels come from strongly magnetised areas, we can obtain accurate results even when the amplitude is up to 1$\times10^{-3}$ of $I_c$. However, the situation is more critical for observations where the dominant magnetic structures are weak and inclined with respect to the solar surface, finding that inaccurate results will start to appear for the inferred magnetic field inclination when signals are on the order of or greater than 5$\times10^{-4}$ of $I_c$. At the same time, we find that the accuracy of the fit is almost identical to the noiseless case when the noise levels are on the order of 1$\times10^{-4}$ of $I_c$. Hence, we only recommend values lower than 5$\times10^{-4}$ of $I_c$ for high-spatial-resolution quiet Sun observations ---such as the one represented here--- if the users of inversion codes aim to reliably interpret the results for the magnetic field vector. We understand that this noise floor seems almost impossible with current observatories unless the spatial information is lost due to the necessary spatial binning to achieve those noise values within reasonable integration times. However, we know, at the same time, that next-generation 4m class telescopes are designed to achieve those levels (although perhaps not at the diffraction limit) with high spatial resolution thanks to the large area of the primary mirror, which can work as a photon collecting bucket at the cost of reducing the spatial resolution.

We also want to emphasise that although there is almost no difference between the results of the inversion without noise and a noise level on the order of 1$\times10^{-4}$ of $I_c$, on many occasions, both inversions results deviate from the original solution. We argue that this is probably a limitation of node-based inversion codes such as SIR and the spectral line selection, which fail to reproduce the stark small-scale variations in height present in the simulated atmospheres. Therefore, in the case of the former, we may, in the future, explore different inversion alternatives \citep[for instance,][]{2019A&A...626A.102A}, while in the case of the latter, we plan to add complementary spectral lines with different sensitivity to the atmospheric parameters, which could help to improve the accuracy of the results. Also, the community has been pushing hard to deliver evermore complex and rich numerical simulations of different solar phenomena, such as those presented in \cite{2017A&A...601A.122D}, and \cite{2019NatAs...3..160C}, and so we plan to use them as laboratory scenarios as well.

Finally, the authors look forward to sharing any data products used in this work, either the atmospheres transformed to optical depth scale or the output spectra at 630~nm, with anyone interested in using them. Please contact the corresponding author with any requests.

\section*{Acknowledgements}

This work has also been supported by the Spanish Ministry of Economy and Competitiveness through the project ESP-2016-77548-C5-1-R, RTI2018-096886-B-C53 and PID2021-125325OB-C51.
This work was supported by Fundação para a Ciência e a Tecnologia (FCT) through the research grants UIDB/04434/2020 and UIDP/04434/2020. This material is based upon work supported by the National Center for Atmospheric Research, which is a major facility sponsored by the National Science Foundation under Cooperative Agreement No. 1852977. This project has received funding from the European Research Council (ERC) under the European Union’s Horizon 2020 research and innovation program (SUNMAG, grant agreement 759548). D. Orozco Suárez also acknowledges financial support through the Ramón y Cajal fellowships.

\bibliographystyle{aa} 
\bibliography{inferb} 

\end{document}